\documentclass[12pt]{article} 

\usepackage{amsmath,amssymb,amsfonts}
\usepackage{psfrag}
\usepackage{color}
\definecolor{darkblue}{rgb}{0.1,0.1,.7}
\usepackage[colorlinks, linkcolor=darkblue, citecolor=darkblue, urlcolor=darkblue, linktocpage]{hyperref} 
\usepackage[square, comma, compress,numbers]{natbib}
\usepackage[]{amsmath}
\usepackage[]{graphicx}
\usepackage[]{latexsym}
\usepackage{geometry}
\usepackage{amscd}
\usepackage[all,cmtip]{xy}
\usepackage{mathrsfs}
\usepackage[margin=10pt,font=small,labelfont=bf]{caption}
\usepackage{dsdshorthand}
\usepackage{simplewick}
\usepackage{changepage}
\numberwithin{equation}{section}

\renewcommand{\be}{\begin{eqnarray}}
\renewcommand{\ee}{\end{eqnarray}}
\newcommand{\bea}{\begin{eqnarray}}
\newcommand{\eea}{\end{eqnarray}}

\def\beq{\begin{equation}} 
\def\eeq{\end{equation}} 
\def\<{\langle}
\def\>{\rangle}

\def\nn{\nonumber} 
 
\def\cO {{\cal O}}

\def\cN {{\cal N}}

\def\CO{{\cal O}}

\begin{document}

\vspace*{-.6in} \thispagestyle{empty}
\begin{flushright}
\end{flushright}
\vspace{.2in} {\Large
\begin{center}
{\bf Covariant Approaches to Superconformal Blocks \\\vspace{.1in}}
\end{center}
}
\vspace{.2in}
\begin{center}
{\bf 
A. Liam Fitzpatrick$^{a}$,
Jared Kaplan$^{b}$,
Zuhair U. Khandker$^{c}$, \\
Daliang Li$^{d}$, 
David Poland$^{d}$,
David Simmons-Duffin$^{e}$} 
\\
\vspace{.2in} 
$^a$ {\it  Stanford Institute for Theoretical Physics, Stanford University, Stanford, CA, 94305}\\
$^b$ {\it  Department of Physics and Astronomy, Johns Hopkins University, Baltimore, MD 21218}\\
$^c$ {\it  Physics Department, Boston University, Boston, MA 02215}\\
$^d$ {\it  Department of Physics, Yale University, New Haven, CT 06520}\\
$^e$ {\it School of Natural Sciences, Institute for Advanced Study, Princeton, New Jersey 08540}
\end{center}

\vspace{.2in}

\begin{abstract}
We develop techniques for computing superconformal blocks in 4d superconformal field theories. First we study the super-Casimir differential equation, deriving simple new expressions for superconformal blocks for 4-point functions containing chiral operators in theories with $\cN$-extended supersymmetry.  We also reproduce these results by extending the ``shadow formalism" of Ferrara, Gatto, Grillo, and Parisi to supersymmetric theories, where superconformal blocks can be represented as superspace integrals of three-point functions multiplied by shadow three-point functions.
 
\end{abstract}

\newpage

\tableofcontents

\newpage


\section{Introduction}
\label{sec:intro}

It is difficult to overstate the importance of conformal field theories (CFTs). They serve as the endpoints of renormalization group flows, they are realized in numerous condensed matter systems at second order phase transitions, they appear to describe consistent theories of quantum gravity through the AdS/CFT correspondence, and they may play some interesting role in physics beyond the Standard Model. While CFTs are in general strongly coupled and difficult to study using the conventional techniques of perturbation theory, it has become apparent in recent years that the conformal bootstrap~\cite{Polyakov:1974gs} approach -- studying the general constraints from symmetries, unitarity, and associativity of the operator product expansion (OPE) -- can be highly successful at making predictions for CFTs in any space-time dimension~\cite{Rattazzi:2008pe,Rychkov:2009ij,Caracciolo:2009bx,Poland:2010wg,Rattazzi:2010gj,Rattazzi:2010yc,Vichi:2011ux,Poland:2011ey,Rychkov:2011et,ElShowk:2012ht,Liendo:2012hy,ElShowk:2012hu,Fitzpatrick:2012yx,Komargodski:2012ek, Beem:2013qxa,Kos:2013tga,Gliozzi:2013ysa,El-Showk:2013nia,Alday:2013opa,Gaiotto:2013nva,Bashkirov:2013vya,Beem:2013sza}. The bootstrap is particularly interesting for supersymmetric theories~\cite{Poland:2010wg,Vichi:2011ux,Poland:2011ey,Beem:2013qxa,Bashkirov:2013vya,Beem:2013sza}, where in addition to having extra symmetry and stronger unitarity constraints, we typically have a greater handle on the space of such theories as well as knowledge of protected aspects of the spectrum. 

An essential ingredient to pursuing the conformal bootstrap is knowing how to decompose 4-point functions into conformal blocks (or conformal partial waves) corresponding to the exchange of primary operators and all of their descendants. In superconformal theories, 4-point functions can be decomposed into superconformal blocks, corresponding to the exchange of superconformal primary operators and all of their superconformal descendants. Past work on superconformal blocks in 4d includes~\cite{Dolan:2001tt,Dolan:2004iy} in $\cN=2,4$ and~\cite{Poland:2010wg,Fortin:2011nq} in $\cN=1$.

In the present paper, we develop two complementary approaches to understanding superconformal blocks, focusing on 4d superconformal field theories. The first approach is to utilize the fact that superconformal blocks can be viewed as eigenfunctions of the super-Casimir differential operator. This approach is particularly straightforward when applied to 4-point functions containing two chiral and two anti-chiral operators, and we derive simple expressions for the corresponding superconformal blocks for any number of supersymmetries $\cN$. However, this approach becomes more cumbersome when applied to more general operators, where the superconformal block can depend on a large number of nilpotent superconformal invariants.

The second approach is to generalize the shadow formalism of Ferrara, Gatto, Grillo, and Parisi~\cite{Ferrara:1972xe,Ferrara:1972ay,Ferrara:1972uq,Ferrara:1973vz}, recently developed further in~\cite{SimmonsDuffin:2012uy}, to superconformal theories. The original idea is that given a CFT operator $\cO(x)$ of dimension $\Delta$ in a $d-$dimensional CFT, one can define a non-local shadow operator $\tilde{\cO}(x)$ with dimension $\tilde{\Delta} = d-\Delta$. Then the integral
\be
\int d^d x \cO(x) |0\>\<0|\tilde{\cO}(x)
\ee
is dimensionless, invariant under conformal transformations, and can be inserted into four-point functions as a projector onto the corresponding conformal block:
\be
\int d^d x \<\phi(x_1) \phi(x_2) \cO(x)\>\<\tilde{\cO}(x)\phi(x_3)\phi(x_4)\> \propto g_{\cO}(x_i) + \textrm{``shadow block"},
\ee
where the shadow block can be easily subtracted off. Similarly, we will show how in a 4d $\cN=1$ SCFT one can take a superconformal primary operator on superspace $\cO(x,\theta,\bar{\theta})$ and define a non-local ``supershadow" operator $\tilde{\cO}(x,\theta,\bar{\theta})$. Then by constructing a superconformally-invariant projector we can project 4-point functions onto simple integral expressions for superconformal blocks. We also apply this method to 4-point functions containing two chiral and two antichiral operators in theories with $\cN$-extended supersymmetry, reproducing the results obtained from the super-Casimir approach. In a companion paper~\cite{Khandker:future} we will further apply it to 4-point functions of real scalar operators in 4d $\cN=1$ theories. 

Both of the two approaches are simplest when described in supertwistor or superembedding space, where the action of the superconformal group $\SU(2,2 | \cN)$ is linearly realized~\cite{Goldberger:2011yp,Goldberger:2012xb,Khandker:2012pa,Siegel:1992ic,Siegel:1994cc,Siegel:2010yd,Siegel:2012di,Kuzenko:2006mv,Kuzenko:2012tb,Maio:2012tx}. 
We review this formalism in Section~\ref{sec:emb}. In Section~\ref{sec:cas} we study the super-Casimir differential equation, focusing on 4-point functions containing chiral operators. In Section~\ref{sec:shadow} we develop an approach to superconformal blocks based on supershadow operators and apply it to the 4-point function containing chiral operators. We conclude in Section~\ref{sec:discussion}.

\section{Superembedding Methods}
\label{sec:emb}
\subsection{Superspace from Supertwistors}

In this section, we review the construction of superspace in terms of objects which transform linearly under superconformal transformations.  We closely follow the discussion of \cite{Siegel:2012di,Khandker:2012pa,Kuzenko:2012tb}, though our notation and conventions are slightly different.  Our construction will enable us to describe certain local operators in a way that makes their superconformal transformation properties manifest. In particular, it will be sufficient to describe general $\cN=1$ superconformal multiplets and some $\cN>1$ multiplets.\footnote{The precise statement is that it can describe multiplets whose superconformal primary is invariant under a nonabelian $R$-symmetry group.  This includes all $\cN=1$ multiplets, since there the $R$-symmetry group is $U(1)$.  However, it does not include many interesting multiplets in theories with $\cN=2,4$, for instance the $\cN=4$ stress-tensor multiplet.  Four point functions of some of these extended SUSY multiplets are discussed in \cite{Arutyunov:2002ff, Dolan:2001tt, Dolan:2004iy}.}  More complicated constructions (e.g. harmonic/analytic/projective superspaces) are required for describing arbitrary multiplets in theories with extended supersymmetry \cite{Galperin:1984av,Galperin:2001uw,Howe:1996rk}. It would be interesting to generalize the techniques in this paper to these spaces.

The building blocks of our construction are supertwistors \cite{Ferber:1977qx},
\be
Z_A =
\begin{pmatrix}
Z_\a\\
Z^{\dot\a}\\
Z_i
\end{pmatrix}\in \C^{4|\cN},
\ee
which have four bosonic components $Z_\a, Z^{\dot\a}$ and $\cN$ fermionic components $Z_i$.  The superconformal group $\SU(2,2|\cN)$ is the subgroup of $\SL(4|\cN)$ that preserves the inner product
\be
\<Z_1,Z_2\> = Z_1^\dag \Omega Z_2,\qquad
\Omega = \begin{pmatrix}
0 & \de^{\dot\b}{}_{\dot\a} & 0\\
\de_{\b}{}^\a & 0 & 0\\
0 & 0 & \de_j{}^{i}
\end{pmatrix}.
\ee
Objects of the form $\bar Z\equiv Z^\dag \Omega$ transform in the dual representation to the supertwistors $Z_A$.  We will call them ``dual supertwistors," with components
\be
\bar Z^A =
\begin{pmatrix}
\bar Z^\a &
\bar Z_{\dot\a} &
\bar Z^i
\end{pmatrix},
\ee
so that $\bar Z_1^A Z_{2A}$ is $\SU(2,2|\cN)$ invariant.

Chiral superspace, with coordinates $(x_+^{\dot\a\a},\th^\a_i)$, is equivalent to the space of two-planes in supertwistor space.  To see why, note that two-planes are spanned by a pair of supertwistors $Z_A^a$, $a=1,2$, subject to a $\GL(2,\C)$ gauge redundancy that acts as a change of basis
\be
Z_A^a &\sim& Z_A^b g_b{}^a,
\qquad
g_b{}^a\in \GL(2,\C).
\label{twistorgaugeequivalence}
\ee 
Here, ``$\sim$" means ``is equivalent to."  Under the action of this $\GL(2,\C)$, a generic pair of supertwistors $Z_A^a$ can be rotated to the form
\be
Z_A^a &=& \begin{pmatrix}
\de_\a{}^a\\
ix_+^{\dot\a a}\\
2\th_i^a
\end{pmatrix}.
\label{eq:poincareslice}
\ee
We refer to this choice of gauge as the ``Poincare slice".

As we see above, the Poincare slice is parameterized by a bosonic vector $x_+^{\dot\a a}$ and $\cN$ fermionic spinors $\th^a_i$, which are the usual coordinates on chiral superspace.\footnote{The tensor $\de_\a{}^a$ in the upper two components of (\ref{eq:poincareslice}) lets us identify the $\GL(2,\C)$ index $a$ with a left-spinor index $\a$.}  The advantage of describing them with the above coset construction is that it makes their transformation law under $\SU(2,2|\cN)$ completely manifest.  If $M\in \SU(2,2|\cN)$ is a superconformal transformation, then we first transform $Z_A^a \to M_A{}^B Z_B^a$.  We then choose a matrix $g\in \GL(2,\C)$ such that $M_A{}^B Z_B^b g_b{}^a$ returns back to the Poincare slice.  The composition of these two transformations defines a map $(x_+,\th)\to(x_+',\th')$ representing the action of $\SU(2,2|\cN)$:
\be
\begin{pmatrix}
\de_\a{}^a\\
ix_+^{\dot\a a}\\
2\th_i^a
\end{pmatrix} = Z_A^a
\quad\to\quad
M_A{}^B Z_B^a 
\quad\sim\quad
 M_A{}^B Z_B^b g_b{}^a = \begin{pmatrix}
\de_\a{}^b\\
ix_+'^{\dot\a a}\\
2\th'^a_i
\end{pmatrix}.
\ee
This precisely reproduces the usual action of the superconformal group on chiral superspace.  

We can similarly describe anti-chiral superspace with the dual twistors $\bar Z^{\dot a A}$.  However, together the objects $Z^a_A,\bar Z^{\dot a A}$ describe 8 real bosonic degrees of freedom (the complex vector $x_+$) and $4\cN$ fermionic degrees of freedom (the spinors $\th_i^\a, \bar\th^{i\dot\a}$).  We need 4 real bosonic constraints to recover the correct degrees of freedom to describe superspace.  Furthermore, these constraints should be superconformally covariant.  The only possibility is
\be
\bar Z^{\dot a A}Z^a_A =0,\qquad a,\dot a=1,2.
\label{ambitwistorrelation}
\ee
In components, this implies
\be
\label{eq:ambitwistorrelationincomponents}
x_+^{\dot\a \a}- (x_+^\dag)^{\dot\a\a}-i4\bar\th^{\dot\a i}\th^\a_i &=& 0,
\ee
which can be solved by writing $x_\pm^{\dot\a \a} = x^{\dot\a\a}\pm 2i\bar\th^{\dot\a i}\th^\a_i$, $x_-=x_+^\dag$, with $x$ real.  In this way, we recover the usual relation between superspace coordiantes $(x,\th,\bar\th)$ and chiral coordinates $(x_+,\th)$.

In what follows, it will often be useful to consider complexified superspace.  For example, correlation functions of local operators can be analytically continued, so they naturally live in complexified superspace.  We will also discuss superspace integration, where one can consider different real contours inside complexified superspace.  In terms of supertwistors, complexification simply means we regard $Z^a_A$ and $\bar Z^{\dot aA}$ as independent, each with their own $\GL(2,\C)$ redundancy
\be
Z_A^a\sim Z_A^b g_b{}^a, \qquad \bar Z^{\dot a A}\sim \bar g^{\dot a}{}_{\dot b}\bar Z^{\dot b A},
\label{eq:gl2redundancies}
\ee
and subject to the (now complex) condition (\ref{ambitwistorrelation}).  The independent supertwistor $Z$ and dual supertwistor $\bar Z$ transform such that the pairing $\bar Z^A Z_A$ is invariant under the complexified superconformal group $\SL(4|\cN)$. 

\subsection{Superembedding Space}

\label{sec:superembeddingspace}

To describe superspace in terms of supertwistors, we were forced to introduce the $\GL(2,\C)\x \GL(2,\C)$ redundancies (\ref{eq:gl2redundancies}).  Physical quantities should be independent of these redundancies, so it's useful to work with objects which transform simply under them.  This motivates the introduction of bitwistors
\be
\label{eq:bitwistordefinitions}
X_{AB}\equiv Z_A^a Z_B^b \e_{ab},\qquad \bar X^{AB}\equiv \bar Z^{\dot a A}\bar Z^{\dot b B}\e_{\dot a\dot b},
\ee
which are well-defined up to rescaling 
\be
\label{eq:bitwistorrescalingredundancy}
(X,\bar X)\sim (\l X, \bar\l \bar X),\qquad \l=\det g, \bar\l=\det\bar g.
\ee
The bitwistor $X$ (and similarly $\bar X$) satisfies the graded antisymmetry relation,\footnote{Note that our definition of $X_{AB}$ differs from that in \cite{Goldberger:2011yp,Khandker:2012pa}, where they satisfy a different antisymmetry condition.}
\be
\label{eq:gradedantisymmetry}
X_{AB}=-(-1)^{p_A p_B}X_{BA},
\qquad
p_A = \left\{
\begin{array}{ll}
0 & \textrm{if }A=\a,\dot\a,\\
1 & \textrm{if }A=i.
\end{array}
\right.
\ee
By construction, $(X,\bar X)$ also satisfy the equations
\be
\label{eq:bitwistororthogonal}
\bar X^{A B} X_{B C} &=& 0,
\ee
and
\be
\label{eq:bitwistortripleantisymmetrize}
X_{[AB}X_{C\}D} = 0, \qquad \bar X^{[AB}\bar X^{C\}D}=0,
\ee
where $[\dots\}$ denotes graded antisymmetrization of indices.

The space in which $(X,\bar X)$ live is called ``superembedding space."  Instead of beginning with supertwistors as we did above, it's possible to describe superspace by working entirely in superembedding space and imposing the equations (\ref{eq:bitwistororthogonal}, \ref{eq:bitwistortripleantisymmetrize}) together with the redundancy (\ref{eq:bitwistorrescalingredundancy}), see for example \cite{Goldberger:2011yp,Khandker:2012pa,Kuzenko:2012tb}.  Both points of view are useful.

Superconformal invariants are given by supertraces of products of
$X$'s and $\bar{X}$'s, for example\footnote{The factor $(-1)^{p_C}$ is necessary to preserve superconformal invariance, since $C$ is contracted from bottom to top, while the superconformally invariant pairing is defined with indices contracted top to bottom.}
\be
\< \bar{2}1\> &\equiv& \bar{X}_{2}^{AB}X_{1BA},\\
\<\bar 4 3 \bar 2 1\>   &\equiv& \bar{X}_4^{AB}X_{3BC}\bar X_2^{CD}X_{1DA} (-1)^{p_C}.
\ee
By construction, these invariants are chiral in unbarred coordinates and antichiral in barred coordinates.

On the Poincare slice, the bitwistors $X_{AB}$ and $\bar X^{AB}$ are given by
\be
X_{AB} &=&
\begin{pmatrix}
\e_{\a\b}
&
-i  (x_+\e)_{\a}{}^{\dot\b}
&
2\th_{j\a}
\\
i(x_+\e)^{\dot\a}{}_\b
&
- x_+^2\e^{\dot\a\dot\b}
&
2i (x_+\th_j)^{\dot\a}
\\
-2\th_{i\b}
&
-2i  (x_+\th_i)^{\dot\b}
&
4\th_i\th_j
\end{pmatrix},
\\
\bar X^{AB} &=&
\begin{pmatrix}
-x_-^2\e^{\a\b}
&
i(\e x_-)^\a{}_{\dot\b}
&
-2i (\bar\th^{ j} x_-)^\a
\\
-i(\e x_-)_{\dot \a}{}^\b
&
\e_{\dot \a \dot \b}
&
 2\bar\th^{j}_{\dot\a}
\\
2i(\bar\th^i x_-)^{\b}
&
-2\bar\th^{i}_{\dot\b}
&
-4\bar\th^{i} \bar\th^{j}
\end{pmatrix}.
\ee
A quantity that will appear frequently is the two-point invariant, which becomes
\begin{equation}
\<\bar 2 1\>=-2\left(x_{2-}-x_{1+}+2i\theta_{1}\sigma\bar{\theta}_{2}\right)^{2}\label{str}\qquad
\textrm{(Poincare slice)}.
\end{equation}

\subsection{Lifting $\cN=1$ Fields to Superembedding Space}
\label{subsec:lifting}

The superembedding space we've constructed is capable of describing all superconformal multiplets in $\cN=1$ theories, and some special multiplets in theories with extended SUSY.  In this section, we briefly summarize the procedure for uplifting fields to superembedding space \cite{Goldberger:2011yp,Goldberger:2012xb,Khandker:2012pa}, focusing on the $\cN=1$ case.\footnote{This is a supersymmetric version of what was presented in \cite{SimmonsDuffin:2012uy}.} 
A four-dimensional $\mathcal{N}=1$ superconformal primary superfield is labeled by its $\SL(2,\mathbb{C})$ Lorentz quantum numbers $(\frac{j}{2},\frac{\bar{j}}{2})$, its scaling dimension $\Delta$, and its $U(1)_R$ charge $R$. It is convenient to summarize these labels as $(\frac{j}{2},\frac{\bar{j}}{2},q,\bar{q})$, where the \emph{superconformal weights} $q,\bar{q}$ are defined by
\begin{equation}
q \equiv \frac{1}{2} \left(\Delta + \frac{3}{2}R\right), \hspace{5mm} \bar{q} \equiv \frac{1}{2} \left(\Delta - \frac{3}{2}R\right) .
\label{q}
\end{equation}

A scalar primary $\phi(x,\theta,\bar{\theta})\sim\left(0,0,q,\bar{q}\right)$ simply gets lifted to a homogeneous scalar $\Phi(X,\bar{X})$ on superembedding space \cite{Goldberger:2011yp},
\begin{equation}
\phi \longrightarrow \Phi,
\end{equation}
\begin{equation}
\Phi:\left(q,\bar{q}\right),
\label{homog}
\end{equation}
where the notation in Eq.~(\ref{homog}) is shorthand for $\Phi(\lambda X,\bar{\lambda}\bar{X})=\lambda^{-q}\bar{\lambda}^{-\bar{q}}\Phi(X,\bar{X})$. 

Handling more general Lorentz representations requires uplifting spinors. A spinor primary $\phi_\alpha\sim\left(\frac{1}{2},0,q,\bar{q}\right)$ gets lifted to a homogeneous dual twistor,
\begin{equation}
\phi_\alpha \longrightarrow \Phi^A,
\end{equation}
\begin{equation}
\Phi^A:\left(q+\frac{1}{2},\bar{q}\right).
\end{equation}
Similarly, a conjugate spinor $\phi^{\dot{\alpha}}\sim(0,\frac{1}{2},q,\bar{q})$ gets lifted to a twistor $\Phi_A$ with homogeneity $\Phi_A: (q,\bar{q}+\frac{1}{2})$. 

The relation between the four-dimensional superfields and their superembedding counterparts is simple,
\begin{eqnarray}
\phi(x,\theta,\bar{\theta}) &=& \left.\Phi(X,\bar{X})\right|_{\mathrm{Poincare}}, \\
\phi_\alpha(x,\theta,\bar{\theta}) &=& \left.\Phi^B(X,\bar{X})X_{B\alpha}\right|_{\mathrm{Poincare}}, \\
\phi^{\dot{\alpha}}(x,\theta,\bar{\theta}) &=& \left.\bar{X}^{\dot{\alpha}B}\Phi_B(X,\bar{X})\right|_{\mathrm{Poincare}},
\end{eqnarray}
where the right-hand side is restricted to the Poincare slice. For operators with spin, we see that contraction with the bitwistors $X,\bar{X}$ projects $\Phi\rightarrow\phi$. In particular, since $\bar{X}^{AB}X_{BC}=0$, there is a gauge-redundancy in the definition of the uplifted field, for instance
\begin{equation}
\Phi^{A}\sim\Phi^{A}+\bar{X}^{AB}\Psi_{B}.
\end{equation}

The spinor case generalizes readily. A generic superfield $\phi_{\alpha_1\cdots\alpha_j}^{\dot{\beta}_1\cdots\dot{\beta}_{\bar{j}}}\sim(\frac{j}{2},\frac{\bar{j}}{2},q,\bar{q})$ lifts to a gauge-redundant multi-twistor $\Phi_{B_1\cdots B_{\bar{j}}}^{\phantom{A_{j}\cdots A_{1}}A_{1}\cdots A_{j}}$ with homogeneity $\Phi:(q+\frac{j}{2},\bar{q}+\frac{\bar{j}}{2})$. It is convenient to introduce index-free notation by using auxiliary twistors $S_{A},\bar{S}^{A}$ to absorb the indices of the superembedding fields. Thus, we define
\begin{equation}
\Phi(X,\bar{X},S,\bar{S})\equiv\bar{S}^{B_{\bar{j}}}\cdots\bar{S}^{B_{1}}\Phi_{B_{1}\cdots B_{\bar{j}}}^{\phantom{A_{n}\cdots A_{1}}A_{1}\cdots A_{j}}S_{A_{j}}\cdots S_{A_{1}} .
\label{IndexFreeField}
\end{equation}
In this language, the gauge-redundancy of $\Phi$ allows us to restrict $S,\bar{S}$ to be transverse and null\footnote{Nullness follows because the transverse conditions can be solved by $S=X\bar{T}$, $\bar{S}=\bar{X}T$ for some $T,\bar{T}$.}
\begin{equation}
\bar{X}S=0,\qquad\qquad\bar{S}X=0,\qquad\qquad\bar{S}S=0.
\label{transverse}
\end{equation} 
Finally, the four-dimensional superfield is recovered by 
\begin{equation}
\phi_{\alpha_1\cdots\alpha_j}^{\dot{\beta}_1\cdots\dot{\beta}_{\bar{j}}} = \left.\frac{1}{j!}\frac{1}{\bar{j}!} \left(\bar{X}\overrightarrow{\partial_{\bar{S}}}\right)^{\dot{\beta}_1}\cdots\left(\bar{X}\overrightarrow{\partial_{\bar{S}}}\right)^{\dot{\beta}_{\bar{j}}} \Phi(X,\bar{X},S,\bar{S}) \left(\overleftarrow{\partial_S} X\right)_{\alpha_1} \cdots \left(\overleftarrow{\partial_S} X\right)_{\alpha_j} \right|_\mathrm{Poincare} .
\label{GeneralProj}
\end{equation}

In what follows, we will be interested primarily in chiral superfields. In superembedding space, chiral fields correspond to holomorphic fields $\Phi(X)$ \cite{Goldberger:2011yp}, i.e. fields that depend only on $X$, not $\bar{X}$, and hence have $\bar{q}=0$. From the projection prescription, Eq.~(\ref{GeneralProj}), it is evident that such a field can only project onto a chiral superfield if $\bar{j}=0$, so that no new $\bar{X}$ dependence is introduced upon projection. This is consistent with the four-dimensional constraint that chiral fields must have $\bar{j}=\bar{q}=0$ \cite{Osborn:1998qu}. Likewise, antichiral fields correspond to antiholomorphic fields $\Phi(\bar{X})$ with $j=q=0$. 

\subsection{Correlation Functions}
\label{subsec:correlators}

Correlators of superembedding fields $\Phi(X,\bar X,S,\bar{S})$, are functions of superconformal invariants built with $S_i$, $\bar{S}_i$, $X_i$, and $\bar{X}_i$ that respect the homogeneity of the constituent fields. In the following discussion we will abbreviate the coordinates $X_i$ and $\bar{X}_i$ simply as $i$, $\bar{i}$ and suppress factors of $(-1)^{p_A}$.

There are two types of such invariants. The first consist of supertraces of coordinates described in Section~\ref{sec:superembeddingspace}, such as $\langle i\bar{j}k\dots\bar{l}\rangle$. There are an infinite number of such supertraces. But for any given number of points, only a finite subset of them are independent. For example, all 3-point invariants built with coordinates are functions of 6 non-vanishing 2-traces: $\langle i\bar{j} \rangle$, where $i,j=1,2,3$ and $i\neq j$.\footnote{For three points, there is one invariant cross-ratio, which can be taken to be $u = \frac{\langle1\bar{2}\rangle \langle2\bar{3}\rangle \langle3\bar{1}\rangle}{\langle2\bar{1}\rangle \langle3\bar{2}\rangle \langle1\bar{3}\rangle}$ and which can appear in three-point correlators of non-chiral fields. We will not need it here.}

Correlation functions of scalar operators are built with such invariants only. In the simple example of the 2-point function of scalar operators $\langle \Phi_1(X_1,\bar X_1) \Phi_2(X_2,\bar X_2) \rangle$, the invariants available are $\langle 1\bar{2} \rangle$ and $\langle 2\bar{1} \rangle$. Imposing homogeneity, one finds that given $\Phi_1\sim(0,0,q,\bar{q})$, the correlator vanishes unless $\Phi_2\sim (0,0,\bar{q},q)$, in which case
\begin{equation}
\< \Phi_1(X_1,\bar X_1) \Phi_2(X_2,\bar X_2) \> = \frac{1}{\langle 1\bar{2}\rangle^{q} \langle 2\bar{1}\rangle^{\bar{q}}} .
\end{equation}

To write down the correlator consisting of operators with non-trivial Lorentz representation, we need invariants that involve auxiliary twistors. In general, these are strings such as  $\bar{S}_p i\bar{j}k\dots\bar{l}S_q$. But not all of them are independent. The following facts facilitate the construction of a non-trivial, independent set of such invariants:
\begin{enumerate}
\item[$\bullet$] By transverseness, Eq.~(\ref{transverse}), $\bar{S}_i$ cannot be contracted with $X_i$, nor $S_i$ with $\bar{X}_i$.
\item[$\bullet$] As a consequence of the graded antisymmetry of $X$, Eq.~(\ref{eq:gradedantisymmetry}), $\bar{S}X\bar{T}=0$ and $S\bar{X}T=0$.
\item[$\bullet$] Eq.~(\ref{eq:bitwistortripleantisymmetrize}) can sometimes be used to reduce long strings of $X$'s and $\bar{X}$'s, for instance $( i\bar{j}i)_{AB} \propto \langle i\bar{j}\rangle i_{AB}$.  
\end{enumerate}

For the 2-point function $\langle \Phi_1(X_1,\bar X_1,S_1,\bar{S}_1) \Phi_2(X_2,\bar X_2,S_2,\bar{S}_2) \rangle$, the considerations above restrict the independent invariants to $\langle 1\bar{2}\rangle$, $\bar{S}_2 1\bar{2} S_1$, and their complex conjugates. Note that the auxiliary twistors only appear in the numerator and their total numbers are restricted by Eq.~(\ref{IndexFreeField}). Imposing homogeneity, one finds that given $\Phi_1\sim (\frac{j}{2},\frac{\bar{j}}{2},q,\bar{q})$, the correlator vanishes unless $\Phi_2\sim (\frac{\bar{j}}{2},\frac{j}{2},\bar{q},q)$, in which case
\begin{equation}
\langle \Phi_1(X_1,\bar X_1,S_1,\bar{S}_1) \Phi_2(X_2,\bar X_2,S_2,\bar{S}_2) \rangle = \frac{(\bar{S}_2 1\bar{2} S_1)^j (\bar{S}_1 2\bar{1} S_2)^{\bar{j}}}{\langle 1\bar{2}\rangle^{q+\frac{3}{2}j} \langle 2\bar{1}\rangle^{\bar{q}+\frac{3}{2}\bar{j}}} .
\end{equation}
The special case that $\Phi_1$ is chiral and $\Phi_2$ is antichiral is given by $\bar{q}=\bar{j}=0$. 

Similar considerations can be used to work out the three-point correlator of a chiral scalar $\Phi\sim(0,0,q_{\Phi},0)$, its antichiral counterpart $\Phi^{\dagger}\sim(0,0,0,q_{\Phi})$, and a real spin-$\ell$ tensor $\mathcal{O}\sim(\frac{\ell}{2},\frac{\ell}{2},q,q)$,
\begin{equation}
\label{3PF}
\langle \Phi(X_{1})\Phi^{\dagger}(\bar{X}_{2})\mathcal{O}(X_0,\bar X_0,S,\bar{S})\rangle =\lambda_{\Phi\Phi^\dagger\cO}\frac{(\bar{S}1\bar{2}S)^{\ell}}{\langle 1\bar{2}\rangle ^{q_{\Phi}-q+\frac{\ell}{2}}\langle 1\bar{0}\rangle ^{q+\frac{\ell}{2}}\langle 0\bar{2}\rangle ^{q+\frac{\ell}{2}}}.
\end{equation}
This correlator will be a starting ingredient for our computation of chiral superconformal blocks via shadow methods in Section~\ref{sec:blocks}.

\section{Superconformal Casimir Approach}
\label{sec:cas}

Conformal partial waves represent the exchange of a definite irreducible representation of the conformal group between pairs of operators.  The conformal Casimir $\mathcal{C}^{(d)}_{\mathcal{N}}$  is an operator that commutes with all conformal generators, so it must have a definite eigenvalue when acting on any single irreducible representation.  Thus the conformal partial waves can be elegantly computed by the eigenvalue problem associated with the conformal Casimir, represented as a differential operator acting on the space of conformally invariant functions.    Let us see how to generalize these ideas to superconformal partial waves.

As a warm-up that is interesting on its own, let us begin by generalizing the two dimensional global or $\SL(2, \C)$ conformal partial waves to superconformal symmetry.  The conformal algebra can be separated into commuting holomorphic and anti-holomorphic parts; the holomorphic part is
\be
[L_n, L_0] = n L_n \ \ \ \mathrm{and} \ \ \ [L_1, L_{-1} ] = 2 L_0 ,
\ee
with $n$ restricted to the values $-1, 0, 1$.  The central charge does not appear in the global conformal algebra.  The holomorphic conformal Casimir  
\be
\mathcal{C}_{0}^{(2)} = L_0^2 - \frac{1}{2} ( L_1 L_{-1} + L_{-1} L_{1} )
\ee
commutes with each of the $L_n$.  The global conformal partial waves in the representation $(h, \bar h)$ of the full $\SL(2, \C)$ have dimension $\Delta = h + \bar h$ and spin $\ell = h - \bar h $; these partial waves  are eigenvectors of the Casimir operator $\mathcal{C}_0^{(2)}$ with eigenvalue $h(h-1)$, and similarly for the anti-holomorphic Casimir.  To make this explicit one computes $\mathcal{C}_0^{(2)}$ as a differential operator acting on the product $\phi(x_1) \phi(x_2)$ within a 4-pt correlator, and then re-writes the result in terms of conformally invariant cross-ratios.

\subsection{$\mathcal{N}=1$ Superconformal Blocks in Two Dimensions}

To generalize to global superconformal symmetry in two dimensions, we extend the holomorphic algebra to include the fermionic generators $G_r$, with $r = \pm 1/2$ and with the (anti-)commutation relations
\be
\{ G_{r}, G_{s} \}  = 2 L_{r+s}  \ \ \ \mathrm{and } \ \ \ [ L_n, G_{\pm \frac{1}{2}} ] =  \left( \frac{n}{2} \mp \frac{1}{2}  \right) G_{\pm \frac{1}{2} + n } .
\ee
This is the global part of the Neveu-Schwarz superconformal sector, where the $r$ indices of the $G_{r}$ take half-integral values \cite{Friedan:1984rv}.  The index $r$ takes integral values in the Ramond sector, but this sector does not have a non-trivial global limit.  Global superconformal primaries are annihilated by both $L_{1}$ and $G_{\frac{1}{2}}$.  The quadratic Casimir
\be
\mathcal{C}_1^{(2)} = L_0^2 - \frac{1}{2} \left( L_1 L_{-1} + L_{-1} L_1 \right ) + \frac{1}{4} \left( G_{+ \frac{1}{2}} G_{-  \frac{1}{2}}   -   G_{-  \frac{1}{2}}  G_{+  \frac{1}{2}} \right) 
\ee
commutes with all the generators.  For a helpful review see e.g. \cite{Friedan:1986rx}.  To compute the superconformal blocks \cite{Belavin:2007zz} we need to represent this algebra as an action on superconformal primaries.  

For this purpose it is sufficient to introduce a single fermionic coordinate $\theta$; superconformal primaries become functions of $(x, \theta)$, where $x$ is a complex coordinate parameterizing the 2-d Euclidean space.
We can represent the action of the algebra on these coordinates as
\be
L_{-1} &=& - \partial_x ,  \\
L_0 &=& -x \partial_x - \frac{1}{2} \theta \partial_\theta , \\
L_1 &=& -x^2 \partial_x - x \theta \partial_\theta ,
\ee  
supplemented by the fermionic generators
\be
G_{-\frac{1}{2}} =  \partial_\theta - \theta \partial_x \ \ \ \mathrm{and } \ \ \ G_{+ \frac{1}{2}} = x \partial_\theta - \theta x \partial_x .
\ee
We will be studying a 4-pt correlator
\be
\mathcal{A}(x_i,\theta_i)  = \langle \phi(x_1, \theta_1)  \phi(x_2, \theta_2)  \phi(x_3, \theta_3)  \phi(x_4, \theta_4)  \rangle 
\ee
and so we need to determine on which superconformal invariants the correlator can depend.  The holomorphic coordinate differences
\be
x_{ij} = x_i - x_j - \theta_1 \theta_2
\ee
are supersymmetric, but not superconformally invariant.  We can construct a pair of superconformal invariants
\be
u = \frac{x_{12} x_{34}}{x_{14} x_{23}} \to \frac{x_1 - x_2 - \theta_1 \theta_2}{x_2} \ \ \ \mathrm{and} \ \ \ v =  \frac{x_{13} x_{24}}{x_{14} x_{23}} \to \frac{x_1}{x_2}
\ee
from the $x_{ij}$, where the latter relations follow when we use a conformal transformation to set $x_3 = 0$ and $x_4 = \infty$.  We can write the correlator or partial wave in this limit as
\be
G(x_1, x_2, \theta_1, \theta_2)  = \frac{1}{ (x_1 - x_2)^{2\Delta_\phi}  (x_3 - x_4)^{2\Delta_\phi} } 
\left[ g_0 \! \left(1 - \frac{x_1}{x_2} \right) + \frac{\theta_1 \theta_2}{x_2} g_\theta \! \left(1 - \frac{x_1}{x_2} \right) \right] .
\ee
In terms of the usual variable $z = 1- \frac{x_1}{x_2}$, the conformal Casimir eigen-equation is
\be
z^2 \left( (1-z) \partial_z^2 - \partial_z  \right) g_0 + \frac{1}{2} z g_\theta &=& q_{h} g_0(z) ,
\\
\left[ z^2  (1-z) \partial_z^2  + z(2-3z) \partial_z  -z + \frac{1}{2} \right] g_\theta + \frac{1}{2} z \left( (1-z) \partial_z^2 - \partial_z  \right) g_0 &=& q_{h} g_\theta(z) ,
\nn
\ee
where $q_h = h (h - \frac{1}{2})$ is the Casimir eigenvalue, where $h$ is the $L_0$ eigenvalue of the primary.  These equations can be solved in terms of hypergeometric functions as
\be
g_0(z) &=& z^{h} {}_2 F_1  \left( h, h, 2h, z \right) , \\
g_\theta(z) &=& h z^{h -1} {}_2 F_1  \left(  h, h, 2h, z\right).
\ee

\subsection{$\mathcal{N}=2$ Superconformal Blocks in Two Dimensions}

The $\mathcal{N}=2$ superconformal algebra has commutation relations
\be
&&[L_m, J_n] = -n J_{m+n}, \qquad \{ G_r^+, G_s^-\} = L_{r+s} + \frac{1}{2} (r-s) J_{r+s} +\frac{c}{6} \left(r^2 - \frac{1}{4} \right) \delta_{s+r,0}  \\
&& \{ G^+_r, G^+_s \} = 0 = \{G^-_r ,G^-_s\} \qquad [L_m, G_r^\pm ] = \left( \frac{m}{2} - r \right) G^\pm_{r+m} , \qquad [J_m, G_r^\pm] = \pm G^\pm_{m+r} \nn
\ee
along with the standard relations for the $L_m$ alone.  Note the addition of the bosonic generator $J_m$, so that we have a new operator $J_0$ in the global limit.
The full Ramond and Neveu-Schwarz algebras are isomorphic in the case of two dimensional $\mathcal{N}=2$ superconformal symmetry.  However, since we are studying the global limit, we will again consider only the NS sector.
The ${\cal N}=1$ generators $G_r$ are $G_r= G_r^+ + G_r^-$.  One can see that dropping $J_m$ for $m\ne 0$ and taking $r,s = \pm \frac{1}{2}$ and $m,n=-1,0,1$, the global algebra closes and the central charge drops out of the commutation relations. The quadratic Casimir is
\be
\mathcal{C}_{2}^{(2)} = L_0^2 - \frac{1}{4} J_0^2 - \frac{1}{2} \{ L_1, L_{-1} \} + \frac{1}{2} [G_+^-, G_-^+]+ \frac{1}{2}[G_+^+,G_-^-] .
\ee
One can represent the $\mathcal{N} = 2$ generators on superspace as
\be
L_{-1} &=& -\partial_x, \\
L_0 &=& - x \partial_x - \frac{1}{2} \theta_1 \partial_{\theta_1} - \frac{1}{2} \theta_2 \partial_{\theta_2}, \\
L_1 &=& x^2 \partial_x - x \theta_1 \partial_{\theta_1} - x \theta_2 \partial_{\theta_2}  , \\
G_-^+ &=& \frac{1}{\sqrt{2}} \partial_{\theta_1} -\frac{1}{\sqrt{2}}  \theta_2 \partial_x , \\
G_-^- &=& -\frac{1}{\sqrt{2}}  \theta_1 \partial_x + \frac{1}{\sqrt{2}} \partial_{\theta_2} , \\
G_+^+ &=& \frac{1}{\sqrt{2}} x \partial_{\theta_1} + \frac{1}{\sqrt{2}} \theta_1 \theta_2 \partial_{\theta_1} -\frac{1}{\sqrt{2}}  \theta_2 x \partial_x , \\
G_+^- &=& \frac{1}{\sqrt{2}} x \partial_{\theta_2} -\frac{1}{\sqrt{2}} \theta_1 \theta_2 \partial_{\theta_2} -\frac{1}{\sqrt{2}}  \theta_1 x \partial_x , \\
J_0 &=& - \theta_1 \partial_{\theta_1} + \theta_2 \partial_{\theta_2}. 
\ee

We can restrict to chiral and anti-chiral fields, meaning fields that are annihilated by $\bar{D}$ and $D$, respectively:
\be
\bar{D}= \partial_{\theta_1} + \theta_2 \partial_x, \nn\\
D=\partial_{\theta_2} + \theta_1 \partial_x .
\ee
Then, a chiral field $\Phi(x,\theta_1, \theta_2)$ depends only on $x-\theta_1 \theta_2$ and $\theta_2$, while an anti-chiral field depends only on $x+\theta_1 \theta_2$ and $\theta_1$. 

To compute the superconformal blocks \cite{Belavin:2012qh} we need to specify the correlator and parameterize it in terms of superconformal invariants.   First, we need to know the supersymmetric distance between two points $(x,\theta_1, \theta_2)$ and $(y, \eta_1, \eta_2)$ in superspace. At linear (quadratic) order in the bosonic (fermionic) components, there are two linearly independent combinations that are invariant under supersymmetric translations and have vanishing R-charge:
\be
\< y \bar{x}\> &\equiv& (y- \eta_1 \eta_2) - (x+\theta_1 \theta_2) - 2 \eta_2 \theta_1, \nn\\
\< x \bar{y}\> &\equiv& (x- \theta_1 \theta_2) - (y+\eta_1 \eta_2) - 2 \theta_2 \eta_1.
\ee
A correlator of generic fields can depend on both of these; however, when chiral or anti-chiral fields are involved, clearly at most one of the above is allowed. Under conformal inversions, individual points transform according to
\be
R&:&  x \rightarrow -\frac{1}{x}, \qquad \theta_1 \rightarrow \frac{\theta_1}x, \qquad \theta_2 \rightarrow \frac{\theta_2}{x}
\ee
Note that under inversions, the chiral position $x-\theta_1 \theta_2$ just becomes the inverse of a chiral position:
\be
x - \theta_1 \theta_2 \stackrel{R}{\rightarrow} -\frac{1}{x} - \frac{\theta_1 \theta_2}{x^2} = -\frac{1}{x - \theta_1 \theta_2}.
\ee
Then, it is easy to see that the chiral-anti-chiral distance $\< x \bar{y}\>$ transforms as
\be
\< x \bar{y}\>  \stackrel{R}{\rightarrow} \frac{ \< x \bar{y}\> }{(x- \theta_1 \theta_2)(y+ \eta_1 \eta_2)}.
\ee
With two chiral and two anti-chiral fields, we can therefore form the invariant $u \equiv \frac{\<1 \bar{2}\> \<3 \bar{4}\>}{\< 1 \bar{4}\>\< 3 \bar{2}\>}$.  We are restricting to purely holomorphic fields, in which case in turns out that this is the only invariant.  This means that our superconformal block depends only on $u$.  Taking the limit where the bosonic component of $\bar{4}$ goes to infinity and all components of $3$ vanish, this simplifies to
\be
u  &\rightarrow& - \left( 1- \frac{x}{y} + \frac{x \eta_1 \eta_2 -2 y \eta_1 \theta_2 +y \theta_1 \theta_2 - \eta_1 \eta_2 \theta_1 \theta_2}{y} \right).
\ee
So, we can act with our Casimir in differential form on the function
\be
g(u) = g_0(z) + \left(  \frac{x \eta_1 \eta_2 -2 y \eta_1 \theta_2 +y \theta_1 \theta_2 }{y} \right) g_2(z) + \frac{x}{y^3} \theta_1 \theta_2 \eta_1 \eta_2 g_4(z)
\ee
where now $z= 1- \frac{x}{y}$, and $g_2(z) = g_0'(z), g_4(z)= g_0''(z) - \frac{g_0'(z)}{1-z}$. We are computing the blocks in the $\phi \times \phi^\dagger$ channel, so we take R-charge to be zero for the internal operator; thus  the eigenvalue of $\mathcal{C}_2^{(2)}$ is $\frac{\Delta^2}{4}$. Acting with the Casimir equation, it is now straightforward to find
\be
g_0(z) = z^h {}_2F_1(h,h, 2h+1, z).
\ee

\subsection{Chiral Blocks in Four Dimensions}

The same methods can be used to compute superconformal partial waves in four dimensions.  The main challenge that one faces in applying this method is the proliferation of independent superconformal invariants.  For this reason, the method is most feasible when applied to superconformal partial waves with chiral and anti-chiral operators.

The $\mathcal{N}=1$ superconformal Casimir operator (in the conventions of Appendix A of \cite{Poland:2010wg}) is
\be
\mathcal{C}_1^{(4)} = \frac{1}{2} M_{\mu \nu} M^{\mu \nu} - D^2 + \frac{3}{4} R^2 + \frac{1}{2} \{ P_\mu, K^\mu \} - \frac{1}{4} [ Q^\alpha, S_\alpha ] - \frac{1}{4}  [ Q^{ \dot \alpha}, S_{ \dot \alpha } ]
\ee
and it takes the eigenvalue 
\be
c_{q, \bar q}^{j, \bar j} = \frac 1 2 j (j+2) + \frac 1 2 \bar j (\bar j + 2) + (q + \bar q)(q + \bar q - 2) - \frac{1}{3} (q - \bar q)^2
\label{eq:n1casimireigenvalue}
\ee
when acting on a state created by a primary operator in the $(\frac j 2, \frac {\bar j} 2)$ Lorentz representation and $(q, \bar q)$ labels the superconformal representation. 

Differential operators representing the superconformal generators are easiest to write down in supertwistor space.  Let us first define generators of $\GL(4|\cN)$ which commute with our $\GL(2)\x\GL(2)$ redundancies and preserve the pairing $\bar Z\. Z$,
\be
L_A{}^B &\equiv& Z^a_A \pdr{}{Z^a_B}-\bar Z^{\dot aB}\pdr{}{\bar Z^{\dot a A}}(-1)^{p_A p_B}.
\ee
The generators of the superconformal group are given by contracting with super-traceless matrices $T^k$, where $k$ indexes the adjoint representation of $\SU(2,2|\cN)$,
\be
(T^k)_B{}^A L_A{}^B(-1)^{p_B}.
\ee

To avoid keeping track of sign factors $(-1)^{p_Ap_B}$, etc. coming from the grading of the components of $Z,\bar Z$, we can make use of the following trick.  Let us pretend that $Z,\bar Z$ are purely bosonic and transform under $\SL(n)$ for some $n$.  $\SL(n)$ invariance will guarantee that the $n$-dependence of our calculations always comes from the trace of the identity matrix $\de^A_A=n$.  In the super case, this trace simply becomes a supertrace.  In other words, we may perform the computation pretending that $Z,\bar Z\in \C^{n\x 2}$, and then set $n=4-\cN$ to recover the answer for the superconformal group $\SU(2,2|\cN)$.

As an example of this trick, let us recover the correct action of the superconformal Casimir on a two-point function.  The Casimir operator for $\SL(n)$ is
\be
C_n=L_A{}^B L_B{}^A-\frac 1 n L_A{}^A L_B{}^B.
\ee
Acting on a two-point function, we get
\be
C^{(1)}_n \frac{1}{\<1\bar 2\>^q \<2\bar 1\>^{\bar q}} &=& \p{2q(2-n+q)+2\bar q(2-n+\bar q) - \frac 4 n (q-\bar q)^2} \frac{1}{\<1\bar 2\>^q \<2\bar 1\>^{\bar q}},
\ee
where the superscript on $C_n^{(1)}$ indicates that the differential operator should act only on $Z_1,\bar Z_1$.  Setting $n=4$, corresponding to $\cN=0$, the quantity in parentheses becomes $(q+\bar q)(q+\bar q - 4)$, which is the correct Casimir eigenvalue for an operator of dimension $q+\bar q$ in a four-dimensional CFT.  Setting $n=3$, corresponding to $\cN=1$, we recover $c^{0,0}_{q,\bar q}$ in (\ref{eq:n1casimireigenvalue}).

Now let us consider a four-point function of chiral and anti-chiral operators
\be
\<\f(X_1)\f^*(\bar X_2)\f(X_3)\f^*(\bar X_4)\>.
\ee
The only superconformal four-point invariants that can be built out of $X_1,\bar X_2, X_3, \bar X_4$ are
\be
\frac{\<1\bar 2 3 \bar 4\>}{\<1\bar 4\>\<3\bar 2\>}\equiv \frac{-1+u+v}{4v}
\qquad
\frac{\<1\bar 2\>\<3\bar 4\>}{\<1\bar 4\>\<3\bar 2\>}\equiv \frac{u}{v},
\ee
where we have defined them in such a way that they reduce to the usual conformal cross-ratios when all the $\th_i, \bar \th_i$ are set to zero.

Acting with the Casimir $C^{(1,2)}_n$ on the ansatz
\be
\<\f(X_1)\f^*(\bar X_2)\f(X_3)\f^*(\bar X_4)\> &=& \frac{1}{\<1\bar 2\>^{\De_\f}\<3\bar 4\>^{\De_\f}} G(u,v),
\ee
we obtain the equation
\be
\label{eq:supercasimireq}
\cD G(u,v)&=&\l G(u,v)\\
\cD &\equiv& ((1-v)^2-u(1+v))\ptl_v v \ptl_v+(1-u+v)u\ptl_u u \ptl_u-2(1+u-v)uv\ptl_v\ptl_u\nn\\
&&-n u \ptl_u+2(n-4)((u-v)u \ptl_u+(1+u-v)v\ptl_v).
\ee
where $\l$ is the Casimir eigenvalue for the exchanged operator.  This equation is closely related to the Casimir equation for a conformal block for scalars $\f_i$ with nonzero $\De_{ij}\equiv \De_i-\De_j$ in a 4d CFT \cite{Dolan:2003hv}.  By relating the differential operators present in the two cases,
one can show that (\ref{eq:supercasimireq}) is solved by
\be
\label{eq:niceexpressionforsuperblock}
G_\cN(u,v) &=& u^{-\cN/2}g_{\De+\cN,\ell}^{\De_{12}=\De_{34}=\cN}(u,v),
\ee
where 
\be
\label{eq:4dconformalblock}
g^{\De_{12},\De_{34}}_{\De,\ell}(u,v) &=& (-1)^\ell\frac{z\bar z}{z-\bar z}(k_{\De+\ell}(z)k_{\De-\ell-2}(\bar z)-(z\leftrightarrow \bar z)) ,\\
k_\b(x) &=& x^{\b/2}{}_2F_1\p{\frac{\b-\De_{12}}2,\frac{\b+\De_{34}}2,\b,x} ,\\
u &=& z \bar{z}, \qquad v \,\,\,=\,\,\, (1-z)(1-\bar{z}),
\ee
is the usual 4d conformal block.

Let us make a few comments about this result.  When $\cN=1$, Eq.~(\ref{eq:niceexpressionforsuperblock}) provides a new  compact expression for the chiral-antichiral block originally derived in \cite{Poland:2010wg}.  Although it is not obvious from the above expression, $G_{\cN=1}$ can be decomposed into a finite sum of $\cN=0$ blocks with $\De_{12}=\De_{34}=0$, as required by the conformal symmetry.

Although our main focus in this paper has been on $\cN=1$ theories, the expression Eq.~(\ref{eq:niceexpressionforsuperblock}) also has meaning when $\cN=2$.  While in general one needs more complicated superspaces to describe CFTs with extended supersymmetry, the superspace defined in Section~\ref{sec:emb} suffices to describe operators which are annihilated by all supersymmetries of one chirality.\footnote{We thank Leonardo Rastelli and Chris Beem for discussions on this point.}  Scalar operators of this type live in so-called $\mathcal{E}_{r(0,0)}$ multiplets \cite{Dolan:2002zh}, and their VEVs parameterize the Coulomb branch of the theory.  In theories with Lagrangian descriptions, examples include $\Tr(\f^k)$, where $\f$ is the adjoint scalar in a $\cN=2$ vector multiplet.  Eq.~(\ref{eq:niceexpressionforsuperblock}) with $\cN=2$ gives the superconformal block for a four point function of such operators and their conjugates.

When $\cN=4$, the constraint that a scalar be invariant under all supersymmetries of one chirality is overly restrictive, and satisfied only by the identity.

\section{Supershadow Approach}
\label{sec:shadow}

In theories whose dynamics respect a symmetry, it is usually fruitful to be able to project transition amplitudes or correlators onto irreducible representations of that symmetry. The shadow operator formalism of Ferrara, Gatto, Grillo, and Parisi \cite{Ferrara:1972xe,Ferrara:1972ay,Ferrara:1972uq,Ferrara:1973vz} was invented to simplify this projection in conformal field theories.  The first observation of this approach is that operators can have non-vanishing two-point function only if they are in representations with the same conformal Casimir, which in terms of the dimension $\Delta$ and Lorentz representation $(j,\bar{j})$ of the primary operator is
\be
C_{\Delta}^{j,\bar{j}} = \Delta (\Delta-4) + C_{j,\bar{j}},
\ee
in $d=4$. Here, $C_{j,\bar{j}} = \frac 1 2 j(j+2) + \frac 1 2 \bar{j}(\bar{j}+2)$ is the Casimir of the Lorentz group.  For a given $C_{\Delta}^{j,\bar{j}}$ and $C_{j,\bar{j}}$, there are therefore two different possible primary operator dimensions in the same selection sector, related by $\Delta \leftrightarrow 4-\Delta$. The representation with primary dimension $4-\Delta$ is referred to as the shadow representation of the primary dimension $\Delta$ representation, and the primary operator with dimension $4-\Delta$ is the shadow operator.  Since both the operator $\CO$ and the shadow operator $\tilde{\CO}$ sit in the same selection sector, either may be used to project onto irreps of $\CO$, but there are certain advantages to using the shadow operator.  Primary among these are that the product $\int d^4 x \CO(x) \tilde{\CO}(x)$ has zero projective weight.  In an appropriately regulated sense, $\< \CO(x) \tilde{\CO}(y)\> \propto \delta^{(4)}(x-y)$, so the shadow operator not only projects onto the irrep of $\CO$, but it also strips out unwanted two-point functions that would arise if we used $\CO$ instead. This fact was used in \cite{Ferrara:1972xe} in order to provide an efficient means of computing the OPE coefficients of descendant operators in terms of those of the primary operators.   The shadow operators are non-local operators; for $j=\bar j=0$ they are
\be
\tilde{\CO}(x) &=& \int d^4 y \frac{1}{(x-y)^{2(4-\Delta)}} \CO(y).
\ee
This manifestly transforms like a primary operator under translations, and by acting with a conformal inversion on $\CO(y)$ and changing integration variables, it is not too hard to see that $\tilde{\CO}$ transforms like a primary operator of dimension $4-\Delta$ under inversions.  Consequently, it transforms like a primary operator of dimension $4-\Delta$ under all conformal transformations.  The two-point function $\< \CO(x) \tilde{\CO}(y)\>$ can easily be regulated and computed by Fourier transforming.

A similar construction is possible and useful in  superconformal theories.  For $\cN=1$, the superconformal Casimir is
\be
C_{q, \bar{q}}^{j,\bar{j}} = (q+\bar{q})(q+\bar{q}-2) - \frac{1}{3} (q-\bar{q})^2 + C_{j,\bar{j}}.
\ee
Thus, in order to satisfy the constraints of R-symmetry on the two-point function and have the same superconformal Casimir, the shadow operator must have $\tilde{q} - \tilde{\bar{q}} = \bar{q}-q$ and $\tilde{q}+\tilde{\bar{q}} = 2-q - \bar{q}$, respectively, so $\tilde{q}=1-q, \tilde{\bar{q}}=1-\bar{q}$.  This is correct dimensionfully for the product $\int d^4 x d^4 \theta \CO(x,\theta, \bar{\theta}) \tilde{\CO}(x,\theta, \bar{\theta})$ to have zero projective weight.  One can construct the shadow operators explicitly as before by using the supersymmetric measure; for $L=0$, now they are\footnote{See \cite{Osborn:1998qu} for details of the conventions adopted here.}
\be
\label{eq:simplesupershadow}
\tilde{\CO}(x,\theta,\bar{\theta}) &=& \int d^4 y d^4 \eta \frac{1}{(x_- - y_+ +4 i \bar{\theta} \eta)^{2(1-\bar{q})}(y_- - x_+ +4 i \bar{\eta} \theta)^{2(1-q)}} \CO^\dag (y,\eta,\bar{\eta}).
\ee


This can be checked by taking a conformal inversion and seeing that $\tilde{\CO}$ transforms the correct way. This follows relatively straightforwardly once one has the transformations under conformal inversions $R$ for $\CO$, the coordinates, and the integration measure, as we discuss in Section~\ref{sec:conventional}.


  One again sees that this is explicitly a non-local operator.  Consequently, when this is used in conjunction with the OPE, one in general has to be careful about the presence of singularities that may arise when the region of integration brings $\CO(y)$ inside the minimal ball surrounding the operators whose OPE is being taken \cite{SimmonsDuffin:2012uy}. Writing the explicit integrals constructing the shadow operators  becomes more involved in standard superspace for operators of higher spin, and it is convenient to pass instead to the super-embedding space.  In Section~\ref{sec:blocks}, we will use the shadow operator formalism together with twistor space to write down integrals that compute the superconformal blocks.

\subsection{Superconformal Integration}

A crucial tool in the shadow formalism is a notion of conformally invariant integration \cite{SimmonsDuffin:2012uy}.  Similarly, here we will need a notion of superconformally invariant integration.  The final answer is simply $\int d^4x d^{4\cN}\th$ with some restrictions on the integrand.  We will arrive at it in two ways: firstly using our realization of superspace in terms of supertwistors, and secondly by a more conventional superspace computation.

\subsubsection{Manifestly Covariant Derivation}

Recall that superspace is given by supertwistors $Z^a_{A}, \bar Z^{\dot a A}$ subject to the condition $\bar Z^{\dot a A}Z^a_A=0$ with a $\GL(2,\C)\x\GL(2,\C)$ gauge redundancy (\ref{eq:gl2redundancies}).  The obvious measure
\be
\prod_{a=1,2} d^{4|\cN}Z^a\prod_{\dot a=1,2} d^{4|\cN}\bar Z^{\dot a}
\ee 
is invariant under $\SL(4|\cN)$, since each term $d^{4|\cN} Z$ transforms with a superdeterminant $\mathrm{sdet}(M)$ under a transformation $Z\to M Z$.

To integrate over superspace itself, we should include the four constraints $\bar Z^{\dot a A}Z^a_A=0$ with a four-dimensional delta function,
\be
\w &\equiv& \prod_{a=1,2} d^{4|\cN}Z^a\prod_{\dot a=1,2} d^{4|\cN}\bar Z^{\dot a} \de^4(\bar Z\. Z).
\ee
Finally, while superconformally invariant, this expression transforms nontrivially under the gauge redundancies (\ref{eq:gl2redundancies}),
\be
\w
&\to&
(\det g)^{2-\cN} (\det \bar g)^{2-\cN} \w.
\ee
Thus, it is only well-defined to integrate $\w$ against a function that transforms oppositely under $\GL(2,\C)\x\GL(2,\C)$:
\be
f(Z g, \bar g \bar Z) &=& (\det g)^{\cN-2} (\det \bar g)^{\cN-2} f(Z,\bar Z).
\label{eq:integrandtransformation}
\ee

For a function $f$ satisfying (\ref{eq:integrandtransformation}), we may define the superconformal integral
\be
\label{eq:superconformalintegral}
\int D[Z,\bar Z] f(Z,\bar Z) &\equiv& \frac{1}{\mathrm{vol}(\GL(2,\C)\x\GL(2,\C))}\int \omega f(Z,\bar Z).
\ee
The integral is gauge-invariant, so it is defined via the Faddeev-Popov procedure.

Passing from the formal definition (\ref{eq:superconformalintegral}) to a more conventional expression is straightforward.  We gauge-fix by choosing $Z$ and $\bar Z$ to lie on the Poincare slice (\ref{eq:poincareslice}) and its dual.  The Faddeev-Popov determinant is trivial, and the argument of the delta function is given by (\ref{eq:ambitwistorrelationincomponents}), so that we have (up to overall constants which we discard)
\be
\int D[Z,\bar Z] f(Z,\bar Z) &=& \int d^4 x_+\, d^4 x_-\, d^{4\cN}\th\, \de^4(x_+-x_--4i\bar\th^i\th_i) f(Z,\bar Z)|_{\textrm{Poincare slice}}\\
&=& \int d^4 x\, d^{4\cN}\th\, f(Z,\bar Z)|_{\textrm{Poincare slice}}.
\ee
We stress that the integral in this simple form is only conformally invariant if $f(Z,\bar Z)$ satisfies the correct homogeneity condition (\ref{eq:integrandtransformation}).

\subsubsection{Conventional Derivation}
\label{sec:conventional}

We can also understand the appropriate superconformally invariant integral in more conventional $\cN=1$ superfield notation.  The integration measure $\int d^4 x d^4 \theta$ is manifestly invariant under translations in superspace, and transforms very simply under dilatations, so the only non-trivial transformation to check is that of conformal inversions.  In general, under a change of variables, the integration measure transforms according to
\be
\int d^4 x d^4 \theta = \int d^4 y d^4 \eta {\rm Ber}^{-1},
\ee
where ${\rm Ber}$ is the Berezinian for the transformation:
\be
 \textrm{Ber} &=& \textrm{sdet} \left( \frac{\partial (y, \eta)}{\partial (x,\theta)} \right) .
 \ee
Since we are interested in conformal inversions, the change in coordinates is
\be
y^{\dot{\alpha} \alpha} &=& \frac{x^{\dot{\alpha} \alpha}}{x^2}, \qquad y_\pm^{\dot{\alpha} \alpha} = \frac{x_{\mp}^{\dot{\alpha} \alpha}}{x_\mp^2} , \qquad \bar{\eta}_{\dot{\alpha}} = - i (x_+^{-1})^{\dot{\alpha}\alpha}  \theta_\alpha, \qquad
\eta^\alpha = i \bar{\theta}_{\dot \alpha} (x_-^{-1})^{\dot{\alpha} \alpha}.
\ee
The computation of $\textrm{Ber}$ for this coordinate change is straightforward but quite long and tedious; the result is
\be
\textrm{Ber} \propto y_+^2 y_-^2.
\ee
Consequently, under a conformal inversion, the coordinates, fields, and integration measure transform according to
\be
\tilde{x}_{\bar{1}2} &\stackrel{R}{\rightarrow}&  x_{1+}^{-1} x_{1 \bar{2}} x_{2-}^{-1}, \\
\CO(y, \eta, \bar{\eta}) & \stackrel{R}{\rightarrow} & 
(y_-')^{2q} (y_+')^{2\bar{q}} \CO(y', \eta', \bar{\eta}'), \\
\int d^4 y d^4 \eta &=& \int \frac{d^4 y' d^4 \eta'}{(y_+')^2(y_-')^2}.
\ee
The shadow field operator is constructed in terms of the original operator $\CO$ through the integral
\be
\tilde{\CO}(x_1,\theta_1,\bar{\theta}_1) &=& \int d^4 x_2 d^4 \theta_2 \frac{1}{(x_{\bar{1}2})^{2(1-\bar{q})}(x_{\bar{2}1})^{2(1-q)}} \bar{\CO}(x_2,\theta_2,\bar{\theta}_2).
\ee
We can take $R\tilde{\CO}R$ by acting on the left and right with $R$ on the right-hand side above.  Crucially, all factors of $(x_{2+}')^2$ and $(x_{2-}')^2$ from the transformation of the operator cancels inside the integrand against the change of the measure and the change of the denominators, to obtain
\be
\tilde{\CO} \stackrel{R}{\rightarrow} (x_{1+}')^{2(1-\bar{q})} (x_{1-}')^{2(1-q)} \int d^4 x_2' d^4 \theta_2' \frac{1}{(x'_{\bar{1}2})^{2(1-\bar{q})}(x'_{\bar{2}1})^{2(1-q)}} \bar{\CO}(x'_2,\theta'_2,\bar{\theta}'_2),
\ee
exactly as necessary for $\tilde{\CO}$ to transform like a superconformal primary operator with $\tilde{q}=1-\bar{q}, \tilde{\bar{q}}=1-q$.

\subsection{Bitwistors, Shadows, and Projectors}
\label{sec:projector}

Working in superembedding space, we can use bitwistors $X,Y$ and the index-free formalism of Section~\ref{subsec:lifting} to define shadow operators and partial-wave projectors in a manifestly-covariant way. For $\mathcal{O}(X,\bar{X},S,\bar{S}) \sim \left( \frac{j}{2},\frac{\bar{j}}{2},q,\bar{q} \right)$, its shadow is given by 
\begin{equation}
\tilde{\mathcal{O}}(X,\bar{X},S,\bar{S})\equiv\int D[Y,\bar{Y}]\frac{1}{\langle X\bar{Y}\rangle ^{2-\cN-q+\frac{j}{2}}\langle \bar{X}Y\rangle ^{2-\cN-\bar{q}+\frac{\bar{j}}{2}}}\bar{\mathcal{O}}(Y,Y\bar{S},\bar{Y}S),
\label{Shadow}
\end{equation}
where $D[Y,\bar{Y}]$, shorthand for $D[Z_Y,\bar{Z}_{\bar{Y}}]$, is the superconformal measure from Eq.~(\ref{eq:superconformalintegral}) and $\bar{\mathcal{O}}\sim ( \frac{\bar{j}}{2},\frac{j}{2})$ is the Lorentz-conjugate of $\mathcal{O}$. Overall, $\tilde{\mathcal{O}}\sim ( \frac{\bar{j}}{2},\frac{j}{2},2-\cN-q,2-\cN-\bar{q} )$ as was noted earlier. Eq.~(\ref{Shadow}) is simply the generalization of Eq.~(\ref{eq:simplesupershadow}) to arbitrary spin, lifted to superembedding space. 

Given a correlation function, the dimensionless projector onto the superconformal multiplet of $\mathcal{O}$ is\footnote{This is the straightforward SUSY generalization of the bosonic embedding-space projector in \cite{SimmonsDuffin:2012uy}.} 
\begin{equation}
\left|\mathcal{O}\right|= \left. \frac{1}{j!^{2}\bar{j}!^{2}}\int D[X,\bar{X}]|\mathcal{O}(X,\bar{X},S,\bar{S})\rangle \left(\overleftarrow{\partial_{S}}X\overrightarrow{\partial_{T}}\right)^{j}\left(\overleftarrow{\partial_{\bar{S}}}\bar{X}\overrightarrow{\partial_{\bar{T}}}\right)^{\bar{j}}\langle \tilde{\mathcal{O}}(X,\bar{X},T,\bar{T})| \hspace{2mm} \right|_M
\label{Projector}
\end{equation}
In particular, for a four-point function $\< \Phi_{1}\Phi_{2}\Phi_{3}\Phi_{4}\>$ the superconformal partial wave $\mathcal{W}_{\mathcal{O}}$ corresponding to $\mathcal{O}$-exchange in the $\left(12\right)\left(34\right)$-channel is given (up to some normalization) by
\begin{equation}
\mathcal{W_{O}}\propto  \left. \< \Phi_{1}\Phi_{2}\left|\mathcal{O}\right|\Phi_{3}\Phi_{4}\> \right|_M.
\label{PWave}
\end{equation}

In the equations above, $\left.\right|_M$ schematically denotes a ``monodromy projection" \cite{SimmonsDuffin:2012uy}. Such a projection should restrict the integral in Eq.~(\ref{Projector}) to only those $X$ compatible with the OPE of the fields $\Phi_i$ appearing in a given correlator. For instance, in Eq.~(\ref{PWave}), the monodromy projection should restrict the integration away from $X_{1,2}$ and $X_{3,4}$ so that the $\Phi_1\times\Phi_2$ and $\Phi_3\times\Phi_4$ OPEs remain valid. Without it, one would have additional ``shadow" partial-wave contributions appearing on the left-hand side of Eq.~(\ref{PWave}). In what follows, we will not need to formulate a supersymmetric definition of monodromy projection, because we will only encounter projections of (non-SUSY) conformal integrals, which have been worked out previously \cite{Dolan:2000ut,SimmonsDuffin:2012uy}. The result is in Eq.~(\ref{Conf Int}). 

From Eqs.~(\ref{Projector}), (\ref{PWave}) we see that in the shadow formalism, the computation of superconformal partial waves boils down to evaluating integrals of the form
\begin{equation}
\mathcal{W}_{\mathcal{O}} \sim  \left. \int D[X,\bar{X}] f(X,\bar{X}) \hspace{2mm} \right|_M
\label{W1}
\end{equation}
where $f(X,\bar{X})$ is essentially a product of a three-point function $\< \Phi_1\Phi_2\mathcal{O} \>$ and a shadow three-point function $\langle \tilde{\mathcal{O}}\Phi_3\Phi_4 \rangle$. Here, we will not attempt to evaluate these integrals in full generality. Rather, we will focus our attention on the case where the superfields $\Phi_i$ in the four-point function, which we refer to as the ``external" fields, are restricted to their lowest component field. The exchanged operator $\mathcal{O}$ remains a full-fledged superfield, so this restricted scenario is still motivated by supersymmetric bootstrap applications. Operationally, setting all $\Phi_i$ to their lowest component is achieved by simply setting their fermionic superspace coordinates $\theta_i,\bar{\theta}_i$ to zero. In the next section, we will show how setting these external thetas to zero in Eq.~(\ref{W1}) can be handled in a manner that preserves manifest (non-SUSY) conformal invariance, reducing the integral in Eq.~(\ref{W1}) to (a possible sum over) known monodromy-projected bosonic conformal integrals.

\subsection{Conformally Covariant Evaluation of Superconformal Integrals}

As explained above, we will be interested in evaluating superconformal integrals with external fermionic coordinates $\th_i,\bar\th_i$ set to zero.  In this subsection, we explain how such integrals reduce to non-SUSY conformal integrals of the type discussed in \cite{SimmonsDuffin:2012uy}.  The result is a compact formula that lets us efficiently evaluate superconformal blocks in terms of conformal blocks. 


\newcommand\cX{\mathcal{X}}
\newcommand\cV{\mathcal{V}}


In our discussion, we will need to distinguish between supertwistors and their bosonic twistor components. For clarity, it will be helpful to modify our notation slightly from that used in previous sections.  Henceforth, we use caligraphic letters $\cZ^a_A, \cX_{AB}$ to denote supertwistors and objects built from them, while reserving roman letters $Z^a_\s, X_{\s\r}$ for restriction to the (bosonic) twistor part, $\s,\r=\a,\dot\a$.  Throughout our computations, we will use the equivalence between antisymmetric bitwistors $X_{\s\r}$ and vectors in the embedding space $X\in \C^6$.

Consider a superconformal integral
\be
I &=& \int D[\cZ,\bar \cZ]g(\cX,\bar \cX),
\ee
where $\cX_{AB},\bar \cX^{AB}$ are bi-supertwistors built from $\cZ,\bar \cZ$ according to (\ref{eq:bitwistordefinitions}), and $g$ is a function of weight $\cN-2$ in both $\cX$ and $\bar \cX$.  We can imagine that $g$ is built from external bi-supertwistors $\cX_i,\bar \cX_i$ together with the integration variables $\cX,\bar \cX$.  

Let us define the fermionic components
\be
\eta^a_I\equiv \cZ_I^a, \quad \bar\eta^{\dot a I}\equiv \bar \cZ^{\dot a I}.
\ee
Where $I=1\dots\cN$ labels the fermionic coordinates of supertwistor space. We will be interested in integrals with the property that when all external Grassmann numbers are set to zero, $g$ is independent of $\eta,\bar \eta$.  The only dependence of the integrand on $\eta,\bar\eta$ is then through the delta function in the measure, so we can immediately integrate over fermionic variables
\be
I &=& \frac{1}{\mathrm{vol}(\GL_2)^2}\int d^8 Z\,d^8 \bar Z d^{2\cN}\eta\, d^{2\cN}\bar\eta\,\de^4(\bar Z^{\dot a \s}Z_\s^a+\bar\eta^{\dot a I}\eta^a_I) g(X,\bar X)\\
&\propto& \frac 1 {\mathrm{vol}(\GL_2)^2} \int d^8 Z\,d^8 \bar Z \p{\p{\e^{ab}\e^{\dot a \dot b}\ptl_{a\dot a}\ptl_{b\dot b}}^\cN\de^4(\bar Z\.Z)}g(X,\bar X).
\label{eq:integralafterfermionicintegration}
\ee

Consider now just the integral over $\bar Z$,
\be
J_h &\equiv& \frac{1}{\mathrm{vol}(\GL_2)}\int d^8 \bar Z \p{\p{\e^{ab}\e^{\dot a \dot b}\ptl_{a\dot a}\ptl_{b\dot b}}^\cN\de^4(\bar Z\.Z)}h(\bar X),
\ee
where $h(\bar X)\equiv g(X,\bar X)$ and for the moment we are pretending that $Z$ and $X$ are constant.  Note that as in Eq.~(\ref{eq:integrandtransformation}), $h$ is homogeneous of degree $\cN-2$.  To proceed, it suffices to compute the above integral on a basis of homogeneous functions of degree $\cN-2$.  As we show in Appendix~\ref{app:feynmanparams}, we can always write $h$ in the form
\be
\label{eq:combineddenominators}
h(\bar X) &=& \sum_P \frac{\G(2-\cN)}{\p{P\.\bar X}^{2-\cN}}
\ee
where $P,\bar X\in \C^6$ are vectors in the embedding space, and the sum over $P$ could be an integral with various weights.  In the case $\cN=2$, one should make sense of this via the replacement\footnote{Although $\log(P\.\bar X)$ transforms via a constant shift under rescalings of $\bar X$, this constant ambiguity will always cancel after taking linear combinations $\sum_P$, so that $h(\bar X)$ is invariant under rescalings, as required.}
\be
\frac{\G(2-\cN)}{(P\.\bar X)^{2-\cN}} &\to& \log(P\.\bar X).
\ee

Thus, let us temporarily replace $h(\bar X)$ with the basis function $\G(2-\cN)(P\.\bar X)^{\cN-2}$ for some $P\in \C^6$.  The answer for the integral is then fixed up to a constant by demanding that it has $SO(4,2)$ invariance, the correct homogeneity in $P$, and also transform appropriately under the $\GL_2$ redundancy acting on $Z$,
\be
\frac{1}{\mathrm{vol}(\GL_2)}\int d^8 \bar Z \p{\e^{ab}\e^{\dot a \dot b}\ptl_{a\dot a}\ptl_{b\dot b}}^\cN\de^4(\bar Z\.Z)\frac{\G(2-\cN)}{\p{P\.\bar X}^{2-\cN}} &\propto& \frac{P^{2\cN}}{(P\.X)^{2+\cN}}\nn\\
&\propto& 
 \left.\ptl_{\bar X}^{2\cN} \frac{1}{\p{P\.\bar X}^{2-\cN}}\right|_{\bar X=X}.
\ee
By linearity, we find 
\be
J_h &\propto& \left.\ptl_{\bar X}^{2\cN} h(\bar X)\right|_{\bar X=X}.
\ee
Substituting this result into (\ref{eq:integralafterfermionicintegration}), we get
\be
I &\propto& \frac{1}{\mathrm{vol}(\GL_2)} \int d^8 Z \left.\ptl_{\bar X}^{2\cN} g(X,\bar X)\right|_{\bar X=X}.
\ee

An integral of this type over a pair of twistors $Z^a$ is equivalent to an integral over the projective null cone in the embedding space
\be
I&\propto& \int D^4 X \left.\ptl_{\bar X}^{2\cN} g(X,\bar X)\right|_{\bar X=X},
\ee
where
\be
\int D^4 X f(X) &\equiv& \frac{1}{\mathrm{vol}(\GL_1)} \int d^6 X \de(X^2) f(X)
\ee
is the conformally invariant integral defined in \cite{SimmonsDuffin:2012uy}.  A simple way to establish the equivalence between these two types of integrals is to show that they agree on a basis of functions with the appropriate homogeneity in $X$, for instance
\be
\frac{1}{\mathrm{vol}(\GL_2)} \int d^8 Z \frac{1}{(P\.X)^4} = \int D^4 X \frac{1}{(P\.X)^4} \propto (P^2)^{-2}
\ee
where $P\in \C^6$ is an embedding space vector.  The $Z$-integral above is evaluated in \cite{Hodges:2010kq}, while the $X$-integral is evaluated in \cite{SimmonsDuffin:2012uy}.  They both equal $(P^2)^{-2}$ (up to numerical constants which can be absorbed into the definition of the integration measure), which is the only possibility consistent with conformal invariance and homogeneity.

To summarize, we have derived
\be
\left.\int D[\cZ,\bar\cZ]g(\cX,\bar\cX)\right|_{\th_i,\bar\th_i=0} &=& \int D^4 X \left.\ptl_{\bar X}^{2\cN} g(X,\bar X)\right|_{\bar X = X}.
\label{IntegralReduction}
\ee
Let us conclude with a brief comment about the meaning of the integrand on the right-hand side.  Since the embedding space vector $\bar X$ is constrained to be null, the operator $\ptl_{\bar X}^{2\cN}$ na\"ively seems ill-defined.  (Since the components of $\bar X$ are not independent, we can't differentiate with respect to each individually.)  However, it happens to be well-defined in the special case we're considering, precisely because $g(X,\bar X)$ is constrained to have degree $\cN-2$ in $\bar X$.

To see why, consider a homogeneous function $h(\bar X)$ with degree $n$ in $\bar X$.  As a function on the null-cone, $h(\bar X)$ is ambiguous up to a shift $h(\bar X)\sim h(\bar X) + \bar X^2 k(\bar X)$, where $k(\bar X)$ is any function of degree $n-2$.  Acting with our differential operator on the ambiguous term, we find
\be
\ptl_{\bar X}^{2\cN}(\bar X^2 k(\bar X))&=& 4\cN(n-\cN+2)\ptl^{2(\cN-1)}k(\bar X) + \bar X^2 \ptl_{\bar X}^{2\cN} k(\bar X)
\ee
where we've used $\bar X\.\ptl_{\bar X} k(\bar X) = (n-2)k(\bar X)$, and $\ptl_{\bar X}^2 \bar X ^2 = 12$, which is twice the dimension of the embedding space.  Precisely when $n=\cN-2$, we have
\be
\ptl_{\bar X}^{2\cN}(\bar X^2 k(\bar X))&=& \bar X^2 \ptl_{\bar X}^{2\cN} k(\bar X)
\ee
Thus, we can set $\bar X^2=0$ either before or after acting with $\ptl_{\bar X}^{2\cN}$, and the result will be consistent.  In other words, when $h(\bar X)$ is restricted to have degree $\cN-2$, the operator $\ptl_{\bar X}^{2\cN}$ maps the ideal generated by $\bar X^2$ to itself, and thus gives a well-defined map on functions on the null-cone.

\subsection{Chiral Blocks in Four Dimensions}
\label{sec:blocks}

\newcommand\cS{\mathcal{S}}

As a simple illustration of the shadow approach, we consider the four-point function of chiral and antichiral superfields in superembedding space,
\be
\< \Phi(\cX_{1})\Phi^{\dagger}(\bar{\cX}_{2})\Phi(\cX_{3})\Phi^{\dagger}(\bar{\cX}_{4})\>,
\label{Chiral4PF}
\ee
where $\Phi\sim\left(0,0,q_{\Phi},0\right)$ and $\Phi^{\dagger}\sim\left(0,0,0,q_{\Phi}\right)$, and compute superconformal blocks corresponding to the exchange of a real spin-$\ell$ operator $\mathcal{O}\sim\left(\frac{\ell}{2},\frac{\ell}{2},q,q\right)$ in the $\Phi\times\Phi^\dagger$ channel. 

The initial ingredients are the three-point function $\langle \Phi\Phi^\dagger\mathcal{O} \rangle$, Eq.~(\ref{3PF}), and its shadow $\< \tilde{\mathcal{O}} \Phi\Phi^\dagger \rangle$, which can be obtained by simply taking $q\rightarrow 2-\mathcal{N}-q$ in Eq.~(\ref{3PF}), i.e.,
\be
\langle \tilde{\mathcal{O}}(\cX_{0},\bar{\cX}_{0},\cT,\bar{\cT})\Phi(\cX_{3})\Phi^{\dagger}(\bar{\cX}_{4})\rangle \propto \frac{(\bar{\cT}3\bar{4}\cT)^{\ell}}{\langle 3\bar{4}\rangle ^{q_{\Phi}-2+\cN+q+\frac{\ell}{2}}\langle 3\bar{0}\rangle ^{2-\cN-q+\frac{\ell}{2}}\langle 0\bar{4}\rangle ^{2-\cN-q+\frac{\ell}{2}}}.
\label{Shadow3PF}
\ee
We will not need to keep track of overall constants. 

The full superconformal partial wave, given by Eqs.~(\ref{Projector}), (\ref{PWave}), is then
\be
\mathcal{W_{O}} \propto \frac{1}{\< 1\bar{2}\> ^{q_{\Phi}-q+\frac{\ell}{2}}\< 3\bar{4}\> ^{q_{\Phi}-2+\cN+q+\frac{\ell}{2}}} \int D[0,\bar{5}]\frac{N_{\ell}}{D_{\ell}},
\label{ChiralW}
\ee
where 
\be
N_{\ell}\equiv\frac{1}{\ell!^{4}}\left(\bar{\cS}1\bar{2}\cS\right)^{\ell}\left(\partial_{\cS}0\partial_{\cT}\right)^{\ell}\left(\partial_{\bar{\cS}}\bar{5}\partial_{\bar{\cT}}\right)^{\ell}\left(\bar{\cT}3\bar{4}\cT\right)^{\ell},
\label{N}
\ee
\be
D_{\ell} \equiv \< 1\bar{5}\> ^{q+\frac{\ell}{2}}\< 0\bar{2}\> ^{q+\frac{\ell}{2}}\< 3\bar{5}\> ^{2-\cN-q+\frac{\ell}{2}}\< 0\bar{4}\> ^{2-\cN-q+\frac{\ell}{2}},
\label{D}
\ee
and we have relabeled $\cX_{\bar{0}}\rightarrow \cX_{\bar{5}}$ to avoid confusion when taking derivatives below.\footnote{The numerator $N_\ell$ can be written as a Gegenbauer polynomial, $N_\ell = \left(-1\right)^{\ell}s^{\frac{\ell}{2}}C_{\ell}^{(1)}(t)$, where $s\equiv\frac{1}{2^{6}}\< 1\bar{5}\> \< 0\bar{2}\> \< 3\bar{5}\> \< 0\bar{4}\> \< 1\bar{2}\> \< 3\bar{4}\>$ and $t\equiv\frac{\< \bar{2}1\bar{5}3\bar{4}0\> }{2\sqrt{s}}$. We will not need to use this fact.} Monodromy projection is understood in all integrals, and we will not write it explicitly. 

We now restrict our attention to the lowest component field of $\Phi$ and $\Phi^\dagger$, setting $\theta_{ext}= 0$.  This amounts to the replacement $\cX_i\to X_i$, where $X_i$ is the top-left $4\x 4$ submatrix of the bi-supertwistor $\cX_i$, along with $\cS\to S, \cT\to T$, where $S$ and $T$ are the twistor parts of the supertwistors $\cS,\cT$.\footnote{Note that the superconformal relations $\bar \cX \cX=0$, $\bar \cS \cS=0$, etc., do not necessarily imply analogous relations among the bosonic twistor components $\bar X X\neq 0, \bar S S\neq 0$.}  As in the previous subsection, we will often think of $X_i$ as a vector in the 6-dimensional embedding space via $X^{\alpha\beta}=\frac{1}{2}X_{m}\Gamma^{m\alpha\beta}$ and $X_{\alpha\beta}=\frac{1}{2}X_{m}\tilde{\Gamma}_{\alpha\beta}^{m}$, where $\Gamma,\tilde{\Gamma}$ are six-dimensional ``sigma"-matrices.  After our replacement, the two-point invariants become $\<i\bar j\> \to -2 X_{ij}\equiv 4 X_i \. X_j$.  Our conventions for embedding space vectors and spinors are those of \cite{SimmonsDuffin:2012uy}.

We then use Eq.~(\ref{IntegralReduction}) to obtain:
\be
\left.\mathcal{W_{O}}\right|_{\theta_{ext}=0}\propto\frac{1}{\left(X_{12}\right)^{q_{\Phi}-q+\frac{\ell}{2}}\left(X_{34}\right)^{q_{\Phi}-2+\cN+q+\frac{\ell}{2}}}\int D^{4}X_{0}\left.\partial_{\bar{5}}^{2\cN}\frac{N_{\ell}}{D_{\ell}}\right|_{\bar{5}=0}.
\label{ChiralW1}
\ee

At this point, our computation boils down to the differentiation in Eq.~(\ref{ChiralW1}), which turns out to be trivial. First, $\partial_{\bar{5}}^{2}N_{\ell}\propto\left(\partial_{\bar{S}}\Gamma^{m}\partial_{\bar{T}}\right)\left(\partial_{\bar{S}}\Gamma_{m}\partial_{\bar{T}}\right)$ $\propto\epsilon^{\alpha\beta\gamma\delta}\partial_{\bar{S}\alpha}\partial_{\bar{T}\beta}\partial_{\bar{S}\gamma}\partial_{\bar{T}\delta}=0$, so
\begin{equation}
\partial_{\bar{5}}^{2}N_{\ell}=0.
\label{Deriv1}
\end{equation}
The mixed derivative $\left(\partial_{\bar{5}}N_{\ell}\right)\cdot\left(\partial_{\bar{5}}D_{\ell}\right)$ contains a term with $\left(\bar{S}1\bar{2}S\right)\left(\partial_{\bar{S}}\bar{1}\partial_{\bar{T}}\right)$ in it and a term with $\left(\partial_{\bar{S}}\bar{3}\partial_{\bar{T}}\right)\left(\bar{T}3\bar{4}T\right)$ in it, both of which vanish since $1\bar{1} = 3\bar{3} = 0$, so
\begin{equation}
\left(\partial_{\bar{5}}N_{\ell}\right)\cdot\left(\partial_{\bar{5}}D_{\ell}\right)=0.
\label{Deriv2}
\end{equation}
Thus the only non-vanishing derivative is
\begin{equation}
\partial_{\bar{5}}^{2}\frac{1}{D_{\ell}}\propto\frac{X_{13}}{X_{1\bar{5}}X_{3\bar{5}}}\frac{1}{D_{\ell}}.
\label{Deriv3}
\end{equation}
The only additional fact \cite{SimmonsDuffin:2012uy} we need is that with $\theta_{ext}=0$,
\begin{equation}
\left.N_{\ell}\right|_{\bar{5}=0} \propto s^{\frac{\ell}{2}}C_{\ell}^{(1)}(t),
\label{Gegen}
\end{equation}
%
where $C_\ell^{(\lambda)}(t)$ are Gegenbauer polynomials and 
\begin{eqnarray}
t &\equiv& \frac{-X_{13}X_{20}X_{40}}{2\sqrt{s}} - \left(1\leftrightarrow2\right) - \left(3\leftrightarrow4\right), \\
s &\equiv& X_{10}X_{20}X_{30}X_{40}X_{12}X_{34}.
\end{eqnarray}
Therefore,
\begin{equation}
\left.\partial_{\bar{5}}^{2\cN}\frac{N_{\ell}}{D_{\ell}}\right|_{\bar{5}=0} \propto \left(\frac{X_{13}}{X_{10}X_{30}}\right)^\cN \frac{\left(X_{12}X_{34}\right)^{\frac{\ell}{2}}C_\ell^{(1)}(t)}{\left(X_{10}X_{20}\right)^q\left(X_{30}X_{40}\right)^{2-\cN-q}}.
\label{Deriv}
\end{equation}
%
Plugging this into Eq.~(\ref{ChiralW1}), we get that 
\begin{equation}
\left.\mathcal{W_{O}}\right|_{\theta_{ext}=0}\propto\frac{\p{X_{13}}^\cN}{\left(X_{12}\right)^{q_{\Phi}-q}\left(X_{34}\right)^{q_{\Phi}-2+\cN+q}} \int D^{4}X_{0}\frac{C_{\ell}^{(1)}(t)}{X_{10}^{\cN+q}X_{20}^{q}X_{30}^{2-q}X_{40}^{2-\cN-q}}.
\label{ChiralW2}
\end{equation}

The integral in Eq.~(\ref{ChiralW2}) is a known (monodromy-projected) conformal integral \cite{SimmonsDuffin:2012uy}:
\begin{equation}
\left.\int D^{4}X_{0}\frac{C_{\ell}^{1}(t_{0})}{X_{10}^{\frac{\Delta+\Delta_{12}}{2}}X_{20}^{\frac{\Delta-\Delta_{12}}{2}}X_{30}^{\frac{\tilde{\Delta}+\Delta_{34}}{2}}X_{40}^{\frac{\tilde{\Delta}-\Delta_{34}}{2}}}\right|_{M} \propto \left(\frac{X_{14}}{X_{13}}\right)^{\frac{\Delta_{34}}{2}}\left(\frac{X_{24}}{X_{14}}\right)^{\frac{\Delta_{12}}{2}}X_{12}^{-\frac{\Delta}{2}}X_{34}^{-\frac{\tilde{\Delta}}{2}}g_{\Delta,\ell}^{\Delta_{i}}(u,v)
\label{Conf Int}
\end{equation}
where $u=\frac{x_{12}^2x_{34}^2}{x_{13}^2x_{24}^2}=z\bar{z}$ and $v=\frac{x_{14}^2x_{23}^2}{x_{13}^2x_{24}^2}=(1-z)(1-\bar{z})$ are the conformal cross-ratios and $g_{\Delta,\ell}^{\Delta_{i}}(u,v)$ are the usual non-SUSY conformal blocks given in Eq.~(\ref{eq:4dconformalblock}).

Using Eq.~(\ref{Conf Int}) to evaluate Eq.~(\ref{ChiralW2}) and dropping the overall constant, our result for the partial wave is 
\begin{equation}
\left.\mathcal{W_{O}}\right|_{\theta_{ext}=0} = \frac{1}{\left(X_{12}\right)^{q_{\Phi}}\left(X_{34}\right)^{q_{\Phi}}} u^{-\frac{\cN}{2}}g_{2q+\cN,\ell}^{\Delta_{12}=\Delta_{34}=\cN}(u,v).
\end{equation}
Peeling off the prefactor $\frac{1}{\left(X_{12}\right)^{q_{\Phi}}\left(X_{34}\right)^{q_{\Phi}}}$ yields the superconformal block for $\mathcal{O}$-exchange,
\begin{equation}
\left.\mathcal{G}_{\Delta,\ell}\right|_{\theta_{ext}=0}=u^{-\frac{\cN}{2}}g_{\Delta+\cN,\ell}^{\Delta_{12}=\Delta_{34}=\cN}(u,v),
\end{equation}
where $\Delta=\Delta_{\mathcal{O}}=2q$. This agrees with our super-Casimir computation, Eq.~(\ref{eq:niceexpressionforsuperblock}).

It is worth emphasizing that the only calculation involved here were the trivial derivatives in Eqs.~(\ref{Deriv1}-\ref{Deriv3}). After performing these embedding-space derivatives, the integral expression for the partial wave simply reduces to a known conformal integral. This is the essence of the shadow approach.

\section{Discussion}
\label{sec:discussion}

Correlation functions in superconformal field theories can be decomposed into partial waves that transform in irreducible representations of the superconformal group.  The superconformal bootstrap program uses these partial waves as atomic ingredients in the bootstrap equation.  The exploration of the SCFT bootstrap is limited by our knowledge of these partial waves in a suitably explicit form.  

In this paper we have presented two formalisms for computing the superconformal partial waves, each generalizing techniques for the computation of conformal partial waves.   In the superconformal Casimir approach, we used the fact that conformal partial waves are eigenfunctions of the quadratic Casimir operator of the superconformal group with an eigenvalue determined by the quantum numbers of the representation.  This approach can be applied in any number of spacetime dimensions, and we gave examples in both $d=2$ and $d=4$.  In the case of chiral and anti-chiral operators, we were able to show that the superconformal blocks can be arranged into a new form equivalent to conformal blocks with quantum numbers shifted by $\cN$.  In particular, we present new results for superconformal blocks of $\mathcal{E}_{r(0,0)}$ multiplets (and their conjugates) in $\cN=2$ theories.  These expressions are immediately applicable to the $\cN=2$ superconformal bootstrap.

However, the super-Casimir approach seems to be of limited utility in the general case due to the proliferation of nilpotent superconformal invariants.  It becomes difficult to solve a differential equation for a function of a large number of independent variables.

Our other approach generalizes the shadow formalism \cite{Ferrara:1972xe,Ferrara:1972ay,Ferrara:1972uq,Ferrara:1973vz,SimmonsDuffin:2012uy} to the superconformal case.  We specialized to $d=4$ in order to write the superembedding space coordinates in terms of supertwistors, which transform naturally under the $SU(2, 2 | \cN )$ group.  This made it possible to write a manifestly invariant projector onto an irreducible representation of the superconformal group using supershadow operators.  The superconformal partial waves were then written as manifestly invariant supertwistor integrals.  In this paper we evaluated a few simple examples involving chiral and anti-chiral primaries.  In a follow-up work \cite{Khandker:future} some of us will use these methods to derive $\cN =1$ superconformal blocks for real scalar operators (including the interesting case of conserved currents \cite{Fortin:2011nq}). These blocks will be essential ingredients for further bootstrap investigations. 

One future direction is to generalize the supershadow approach to incorporate more complicated superspaces describing other $\cN>1$ multiplets. In particular the harmonic superspace for $\cN=2$ would be an interesting starting point. 
\section*{Acknowledgements}

We thank D. Skinner for essential discussions at the initial stages of this project.  We are also grateful to C. Beem, K. Intriligator, L. Rastelli, and A. Stergiou for discussions. We additionally thank the organizers of the ``Back to the Bootstrap 2" workshop at the Perimeter Institute and the ``Back to the Bootstrap 3" workshop at CERN for facilitating discussions related to this work. DSD was supported by DOE grant DE-SC0009988.  JK was supported in part by the National Science Foundation grant PHY-0756174.   ALF was partially supported by ERC grant BSMOXFORD no. 228169. ZUK is supported by DOE grant DEFG02-01ER-40676. 

\newpage
\appendix

\section{A Basis for Homogeneous Functions}
\label{app:feynmanparams}

Consider a homogeneous function $h(X)$ of degree $-d$ in a vector $X$.  We claim that $h$ can be written as a linear combination of functions of the form $(P\.X)^{-d}$.  For the purposes of this work, it suffices to consider products
\be
h(X) &=& \prod_i (A_i\.X)^{-a_i},
\ee
where $\sum_i a_i = d$.  Feynman parameters don't work if any of the $a_i$ are negative.  To address this, choose integers $k_i$ such that $a_i+k_i>0$.  We may now safely write
\be
h(X) &=& \prod_i \frac{(A_i\.X)^{k_i}}{(A_i\.X)^{a_i+k_i}}\nn\\
&=& \frac{(-1)^{\sum_i k_i}}{\prod_i \Gamma(a_i+k_i)}\int \de(1-\sum_i t_i) \prod_i \frac{dt_i}{t_i} t_i^{a_i+k_i} \ptl_{t_i}^{k_i} \frac{\G(d)}{(P(t_i)\.X)^{d}}.
\label{eq:generalizedfeynmanparam}
\ee
where $P(t_i) \equiv \sum_i t_i A_i$.  When $d=0$, we can replace
\be
\frac{\G(d)}{(P(t_i)\.X)^d} &\to& 
\log (P(t_i)\.X)
\ee
and Eq.~(\ref{eq:generalizedfeynmanparam}) remains true.

\bibliography{Biblio}{}
\bibliographystyle{utphys}

\end{document}